\documentclass[journal]{IEEEtran}

\usepackage[lofdepth,lotdepth]{subfig}
\usepackage{graphicx}
\usepackage{algorithm}
\usepackage{algpseudocode}
\usepackage{cite}
\usepackage{amssymb}
\usepackage{gensymb}
\usepackage{hyperref}

\usepackage[latin1]{inputenc}
\usepackage{tikz}
\usetikzlibrary{shapes,arrows,positioning,calc}

\usepackage{color}
\usepackage[normalem]{ulem}

\tikzstyle{startstop} = [rectangle, rounded corners, minimum width=3cm, minimum height=1cm,text centered, draw=black, fill=red!30]
\tikzstyle{io} = [trapezium, trapezium left angle=70, trapezium right angle=110, minimum width=3cm, minimum height=1cm, text centered, draw=black, fill=blue!30]
\tikzstyle{process} = [rectangle, minimum width=3cm, minimum height=1cm, align=left, draw=black, fill=orange!30]
\tikzstyle{empty} = [rectangle, align=left]
\tikzstyle{decision} = [diamond, minimum width=3cm, minimum height=1cm, align=center, draw=black, fill=green!30]
\tikzstyle{arrow} = [thick,->,>=stealth]

\begin{document}
%
\title{AC Loss Modelling in Superconducting Coils and Motors with Parallel Tapes as Conductor}

\author{Enric Pardo \IEEEmembership{Senior Member,~IEEE}, Francesco Grilli, Yingzhen Liu, Simon Wolft\"adler, Thomas Reis 
\thanks{Enric Pardo is with the Institute of Electrical Engineering, Slovak Academy of Sciences, Dubravska 9, 84104 Bratislava, Slovakia.}
\thanks{Francesco Grilli and Yingzhen Liu are with Karlsruhe Institute of Technology, Germany.}
\thanks{Simon Wolftaedler and Thomas Reis are with OSWALD Elektromotoren GmbH, Miltenberg, Germany.}
\thanks{Funded by the  European Commission Grant No 723119 (ASuMED). IEE also acknowledges the financial support of the Grant Agency of the Ministry of Education of the Slovak Republic and the Slovak Academy of Sciences (VEGA) under contract No. 2/0097/18, the Slovak Research and Development Agency under the contract No. APVV-14-0438. The calculations from IEE were performed in the Computing Centre of the Slovak Academy of Sciences using the supercomputing infrastructure acquired in project ITMS 26230120002 and 26210120002 (Slovak infrastructure for high-performance computing) supported by the Research \& Development Operational Programme funded by the ERDF.}
\thanks{\color{blue} This article has been published in IEEE Transactions on Applied Superconductivity, and hence it is subject to copyright. Cite as: E Pardo et al., IEEE Trans. on Appl. Supercond., vol. 29, iss. 5, a.n. 5202505, 2019.}
\thanks{\color{blue} Article DOI: 10.1109/TASC.2019.2899148}
}




\maketitle

\begin{abstract}
Many superconducting power applications contain {REBa$_2$Cu$_3$O$_{7-x}$ (or REBCO)} coils{, where RE stands for ``rare earth"}. Superconducting motors are very interesting for mobile applications like aviation, thanks to their high power per unit weight. In order to reduce the inductance, multi-tape conductors are often used in windings but parallel tapes present coupling AC loss. In this article, we analyze the AC loss of single coils and a full stator of a motor made of windings of parallel tapes, taking several coupling situations into account. Our computer modeling method combines self-programmed and commercial finite-element methods to calculate the current density and AC loss. We found that isolating the tapes along the length provides almost the same AC loss reduction as that obtained by full transposition. This reveals that there are still many unexploited strategies to reduce the AC loss in multi-tape conductors.
\end{abstract}



\section{Introduction}

REBCO coated conductors are very promising materials for power applications thanks to their high opearation temperature and good performance in high magnetic fields compared to other superconductors \cite{wolsky13IEA,martucciello15SST}. Many superconducting power applications contain coils, such as transformers, motors, generators and superconducting magnetic energy storage (SMES). 

Superconducting motors are particularly interesting for mobile applications, such as sea transport and aviation \cite{gieras09PET,haran17SST}. Superconducting motors are a key enabler for distributed electric propulsion in aircrafts. Hybrid-electric distributed propulsion can reduce CO$_2$ and NO$_x$ emissions by 75 \%, fuel burnt by 70 \% and noise by 71db \cite{haran17SST,ASuMED}. {Superconducting motors are very promising to achieve the high required specific power, being between 10 and 15 kW/kg according to Ref. \cite{delhaye15talk}.}


In order to reduce the inductance, multi-tape conductors are often used in the windings. The simplest solution is to stack tapes in parallel. However, for this case coupling effects appear due to the magnetic field component paralel to the tape surface \cite{acreview}. A way to reduce the coupling effects is to transpose the tapes in complex wire configurations, such as ROEBEL \cite{roebelreview}, CORC \cite{weiss17SST,vanderlaan18SST,vojenciak15SST,souc17SST}, and twisted stack cables \cite{takayasu16IES,bykovsky16IES}. However, these configurationis increase the conductor cost and, in some cases, they decrease the engineering current density. Although CORC and twisted stacks provide the necessary transposition for striated tapes, their application to power applications is very challenging. The reason is that power frequencies are relatively high (above 50 Hz) compared to proposed solutions for magnets, with characteristic frequencies below 1 mHz \cite{amemiya18SST}.

Although there are many works studying the AC loss of single pancake coils \cite{grilli07SST,clem07SSTb,pancaketheo,pardo12SSTb,zermeno11IES,ainslie11SST,acreview}, very few take parallel tapes into account \cite{grinenko12SST,grinenko13SST}. For the latter, coils with only up to 4 tapes are measured.

In this article, we analyze the AC loss in coils made of cables with up to 16 untwisted parallel tapes for several coupling limits. We show by numerical modeling that using untransposed electrically isolated tapes, only soldered together at the terminals, reduces the AC loss practically to the transposed case. We also show that this approach is useful to reduce the AC loss in the stator windings of real motors. For this, we used the geometry of an existing superconducting motor \cite{oswald12PhP}. Preliminary results of this article have been presented in \cite{motor17EUCAS,motor18HTSmod}.


\section{Numerical method}

{We assume a power-law relation between the electric field{, $E$,} and the current density{, $J$, as $E(J)=E_c ( {|J|}/{J_c} )^{n-1} ({J}/{J_c})$, where $E_c=10^{-4}$ V/m and $n$ is the power-law exponent. Here, we used $n=$30, which provides results close to the critical-state model limit ($n\to\infty$ \cite{acreview}}).

In this work, we assume {uniform} $J_c$. Previous work on transformers \cite{transformer} {showed} that models with this assumption can predict the AC loss with a 10-20 \% accuracy, if $J_c$ is taken as the coil average $J_c$ (average $J_c$ among all coil turns). More accurate modeling could be done {in the future} by taking an anisotropic magnetic-field dependent $J_c$ into account.}

For stand-alone coils, we model the superconductor by the Minimum Electro-Magnetic Entropy Production (MEMEP) \cite{pardo15SST} method. For the motor, we combine MEMEP with conventional static Finite-Element Methods (FEM). With this combination, we can calculate the critical current penetration process, local AC loss, and {total loss} in the whole superconducting stator. In this article, we only take cross-sectional two-dimensional (2D) modeling into account.

The calculations made by MEMEP use a numerical tool programmed in C++, which calculates only the response of the superconductor. We can take the interaction with the rest of the motor into account by means of an applied vector potential $A_a(x,y)$, corresponding to the vector potential created by the iron parts and the rotor. We calculate $A_a(x,y)$ by means of COMSOL in the $A$ formulation. For this purpose, we assume uniform current density, $J$, in the superconducting coils and {we} later find $J(x,y)$ by MEMEP, where $x$ and $y$ are the Cartesian space coordinates. In particular, we evaluate $A_a (x,y)$ at rectangles containing the superconducting parts for a given number of time instants, which cover at least one full electric cycle. In this article, we used up to 72 time steps per cycle.

Since we solve the ferromagnetic material by assuming constant $J$ in the coils, this method does not take into account the interaction between magnetization currents {(current density contribution differing from uniform current density)} in the superconductor and the ferromagnetic materials. We also assume that the magnetization currents do not modify the rotor properties. The assumed model is valid when the separation between the superconductor and the iron parts (or the superconductor in the rotor) is larger than the tape width. {This is because} at distances larger than the tape width, the magnetic field from magnetization currents is much smaller than the transport contribution (or contribution from uniform $J$). {The reason is that the dipolar and higher multipole contributions from the magnetization currents decrease at least as $1/r^2$, being $r$ the distance; while the monopolar component, which is only present in the transport contribution, decreases as only $1/r$. Our model has been checked by comparing the AC loss results to those obtained by FEM in $T-A$ formulation taking the full interaction into account. We obtained a typical deviation of 5.6 \% with a maximum of 10 \%, being the results for the model in this article always higher \cite{motorTA}.} In any case, {the method in this article} does take into account:
\begin{itemize}
\item The interaction of magnetization currents between all superconducting tapes in
all coils.
\item The effect of iron and rotor on the superconductor response, in particular $J(x,y)$
and AC loss.
\end{itemize}
Although the coil sections of the motor in this article are relatively separated from each other (see Fig. \ref{fJmot}), we take the interaction of all superconducting parts into account, consisting on 1248 tapes in the cross-section.

In this article, we use a discretization where there is only one cell in the tape thickness. By doing this, we implicitly take the thin film approximation, since all electromagnetic quantities are averaged across the tape thickness. However, the model enables to use several elements on the thickness.


\begin{figure}[tbp]
\centering
{\includegraphics[trim=0 0 0 0,clip,width=8 cm]{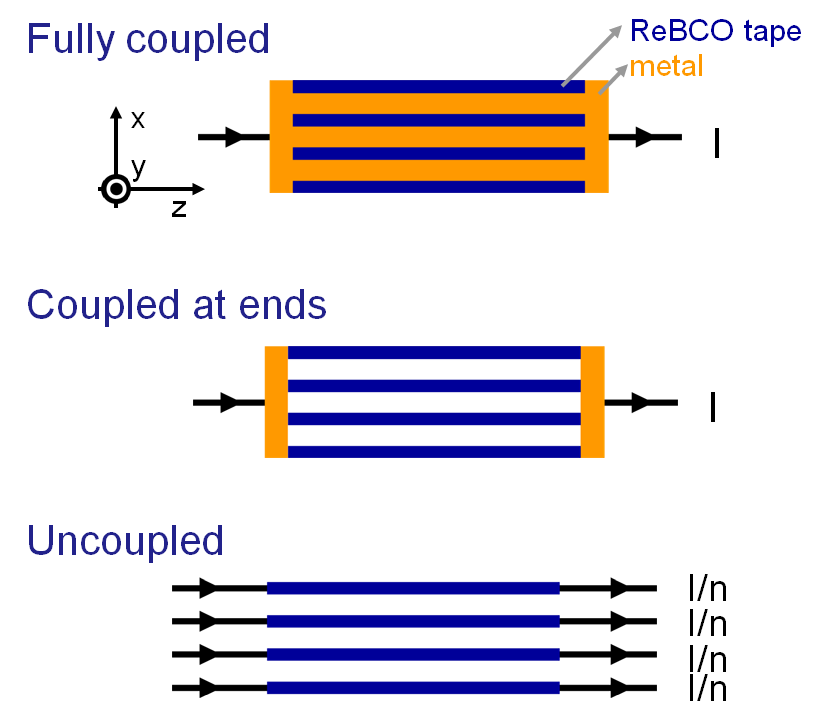}}
\caption{The three studied coupling situations above result in different AC loss in the superconductor. $I$ is the cable transport current and $n$ is the number of tapes in the conductor.}
\label{fsketch}
\end{figure}

We take three different coupling situations into account: fully coupled, coupled at ends, and uncoupled (Fig. \ref{fsketch}). The fully coupled case corresponds to having a very good conductor connecting all superconducting tapes along its length. The coupled at ends scenario is when this soldering is only at the terminals. Finally, the uncoupled case is the ideal situation that the transport current is fully shared between all tapes, being equivalent to having a current source for each tape or perfect transposition. For all cases, we take into account the effect in the AC loss created by the superconductor, which can vary by orders of magnitude. We implement these three scenarios by properly choosing the current constrains in each cable:
\begin{itemize}
\item Uncoupled: same net currents in all tapes.
\item Coupled: free current distribution among all tapes in a cable.
\item Coupled-at-ends: a given tape in a turn cross-section has the same net current in all turns. For instance, the leftmost tape in the left half-section of Fig. \ref{fJ1} has the same net current as all left-most tapes in the left half-section and all right-most tapes in the right half-section.
\end{itemize}

We do not take the direct loss contribution from the metal parts into account. This assumption can be done when the length of the superconducting cable is high or the resistance between tapes is very low (high conductance per unit length for the coupled case), so that the frequency is much above the peak of the coupling loss. The AC loss at the connections for an application can be minimized by reducing the number of terminals.


\section{Single coil}

We assume racetrack coils much longer than its width made of 4 turns of 16 tapes in parallel. Each tape is 4 mm in width and has a superconducting layer of around 2 $\mu$m (value taken in modelling). The separation between superconducting layers in tapes of the same cable is 158 $\mu$m, being 358 $\mu$m for neighboring tapes in different cables. The coil inner bore is 30 mm.

\begin{figure}[tbp]
\centering
{\includegraphics[trim=0 -30 0 0,clip,width=8 cm]{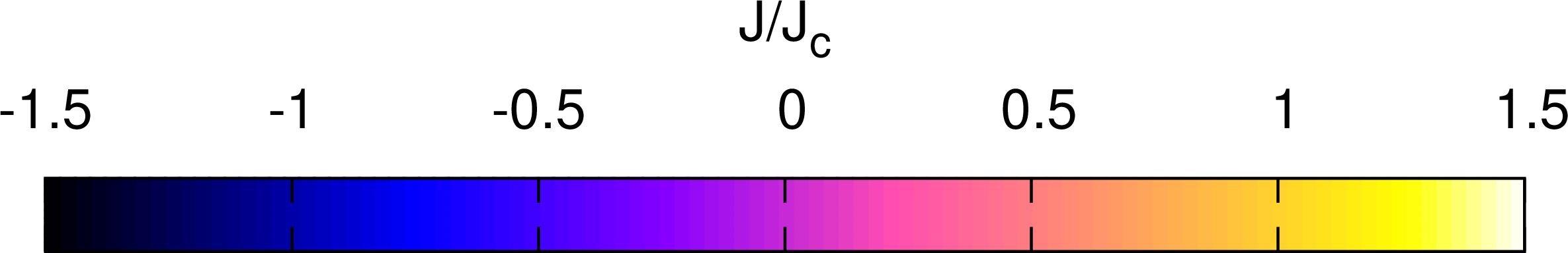}} \\
{\includegraphics[trim=0 170 0 0,clip,width=9 cm]{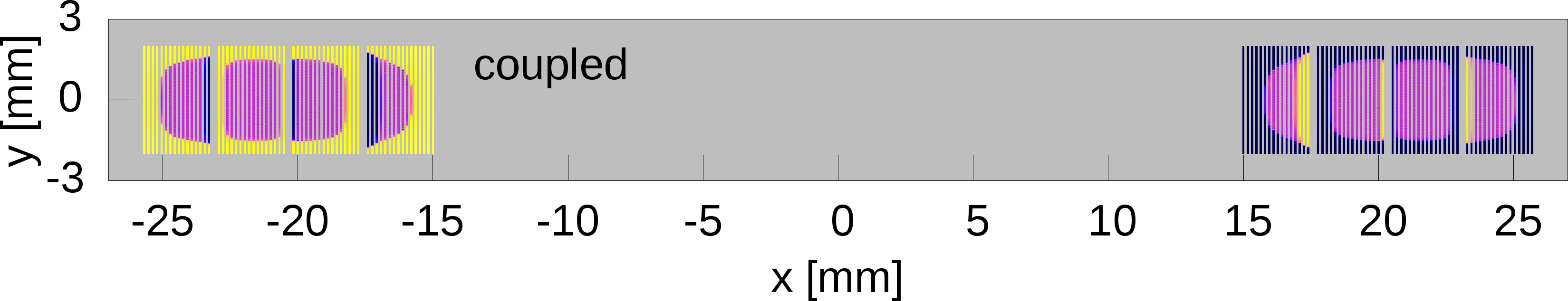}} \\
{\includegraphics[trim=0 170 0 -20,clip,width=9 cm]{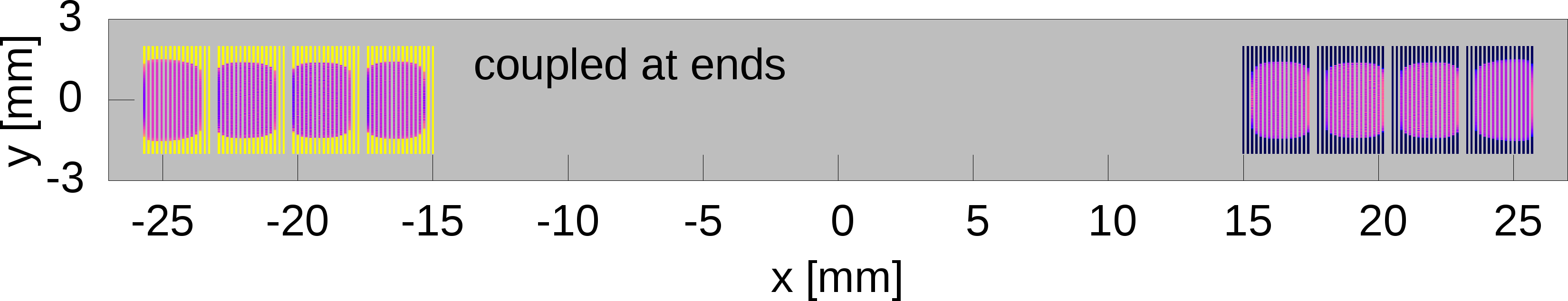}} \\
{\includegraphics[trim=0 0 0 -20,clip,width=9 cm]{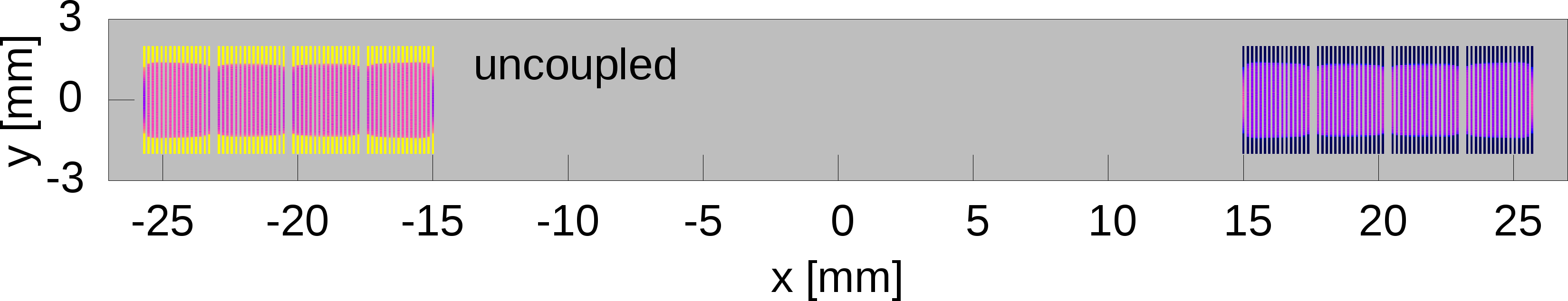}}
\caption{In a stand-alone coil, the coupled case presents the highest $J_c$ penetration. Case for 16 tapes in parallel and peak of the alternating current. {The current amplitude is 50 \% of the critical current and the frequency is 500 Hz.}}
\label{fJ1}
\end{figure}

The coil presents current density with opposite sign from the net transport current in the coupled case, evidencing large magnetization currents (Fig. \ref{fJ1}). For the uncoupled case, current penetration appears almost only from the top and bottom of the winding. The coupled-at-ends situation is {qualitatively} in between but it does not present currents with opposite sign. The reason is that the net magnetic flux between tapes in the cable nearly vanishes in each half-section of the coil. For the limit of a coil width much lower than the coil bore, this cancellation is exact.

\begin{figure}[tbp]
\centering
{\includegraphics[trim=0 0 0 0,clip,width=8.5 cm]{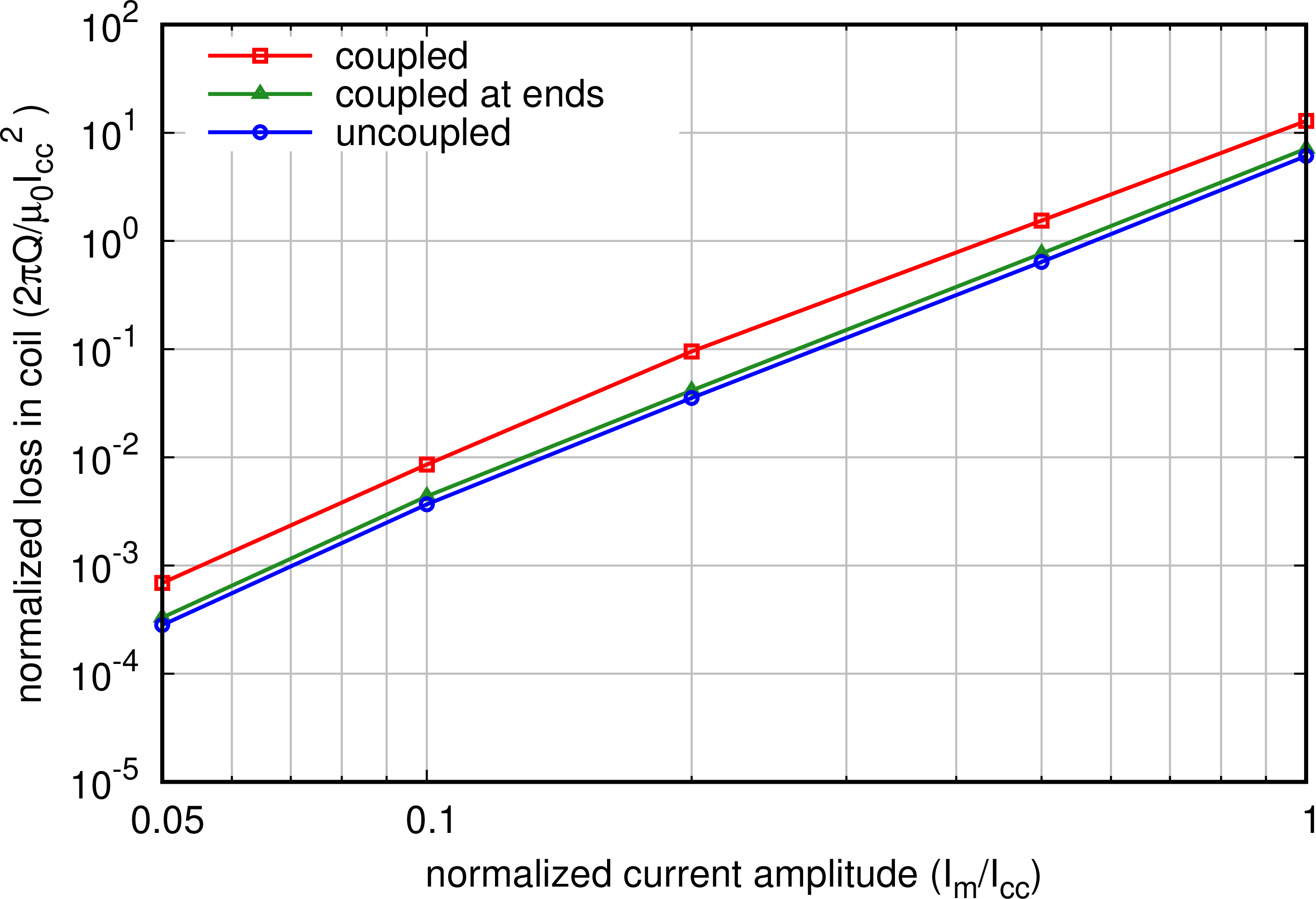}} 
\caption{In a stand-alone coil, the AC loss for the coupled-at-ends case is almost the same as the uncoupled situation, representing ideal transposition. Above, $Q$ is the AC loss per cycle and coil length, $I_m$ is the current amplitude in the cable, and $I_{cc}$ is the cable critical current.}
\label{fQ1}
\end{figure}

As a consequence of this partial cancellation of the flux between tapes in the cables, the AC loss for the coupled-at-ends scenario is practically the same as the {uncoupled} situation (see Fig. \ref{fQ1}). The uncoupled case would require to either transpose the conductor or use separate power sources for each tape. Then, the coupled-at-ends situation is the preferred kind of coupling, since it is much simpler to achieve in practice than the uncoupled situation. Since we {use a high power-law exponent in the $E(J)$ relation, the results are approximately the same as for the critical-state model assumption ($n\to\infty$). Then,} the normalized AC loss of {Fig.} 6 is useful to {estimate} the AC loss for any given tape critical current.

\section{Motor}


In this section, we study the AC loss in the stator of a particular superconducting motor. The superconducting torque motor (SUTOR) is a synchronous motor with permanent magnets \cite{oswald12PhP}. Detailed data is presented in table \ref{t.SUTOR}. High temperature superconductors are used in the stator winding system combined with rare-earth permanent magnets on the rotor. The cryogenics of the system is designed as a closed loop cooling circuit, which uses helium gas to cool the stator coils. These coils are placed in single cryostats made of polyetherimide, enabling to operate the remaining motor sections (stator-, rotor- iron and permanent magnets) at ambient temperature.

\begin{table}[tpb]
\begin{center}
\caption{{Parameters of the superconducting SUTOR motor. Cross-section in {Fig.} \ref{fgeomot}.}}
\label{t.SUTOR}
\begin{tabular}{ll}
\hline
\hline
\# teeth & 12 \\
\# pole pairs & 4 \\
\# slots per pole and phase & 0.5 \\
\# coils & 12 \\
\# turns per coil & 52 (4 in this paper) \\
\# superconducting tapes in parallel & 1 (13 in this paper) \\
Active length & 240 mm \\
Outer diameter & 288 mm \\
Nominal current per coil (rms) & 32.5 A (442.5 A in this paper) \\
Current in coil section (rms) & 1690 A$\cdot$turns \\
Nominal mechanical speed & 665 1/min \\
Nominal mechanical torque & 575 Nm \\
Nominal electrical frequency & 44.333 Hz \\
Permanent magnet material & NdFeBr ($B_r$=1.2 T, $\mu_r$=1.05) \\
Stator and rotor material & M270-35A (DIN EN 10106) \\
Superconducting material & 2G REBCO tape \\
& ($I_c$=130 A at self-field) \\
Cooling method & Closed loop system with helium gas. \\
Cryostat & Single cryostat design for each coil  \\
& (polyetherimide). \\
\hline
\hline
\end{tabular}
\end{center}
\end{table}

In this article, we use a large number of tapes in parallel (13 tapes) in order to study the effect of coupling in a realistic motor environment.

\begin{figure}[tbp]
\centering
{\includegraphics[trim=0 0 0 0,clip,width=8.5 cm]{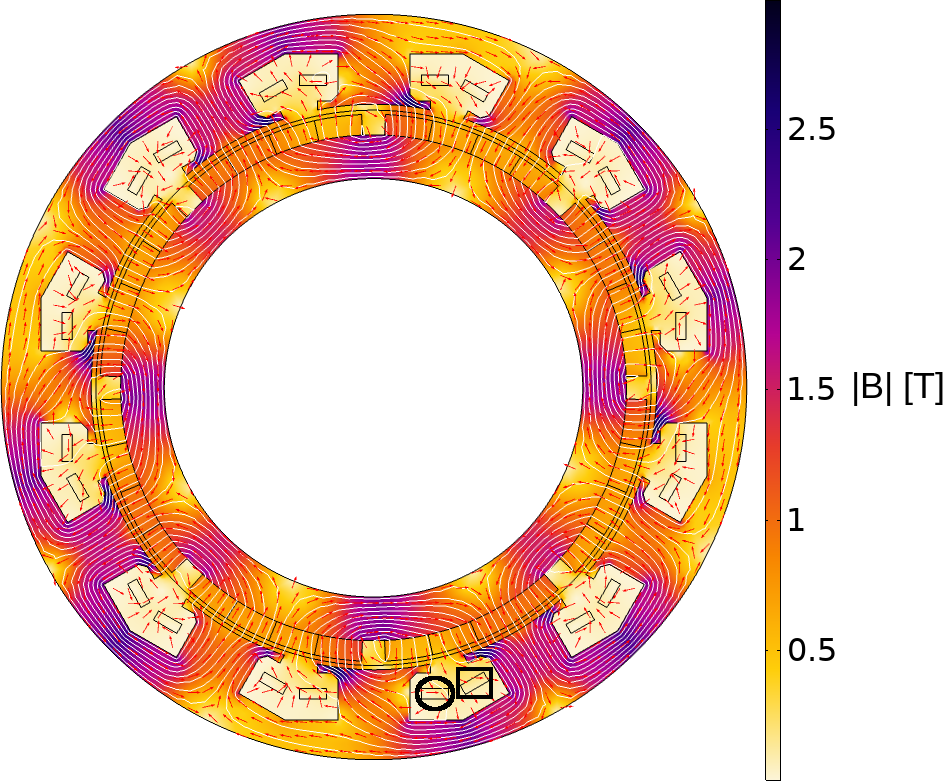}} 
\caption{The superconducting stator enhances the torque and torque per weight of the SUTOR motor \cite{oswald12PhP}. Parameters in table \ref{t.SUTOR}. Circled and framed cross-sections detailed in Fig. \ref{fJmot}.}
\label{fgeomot}
\end{figure}

The magnetic flux density at the initial phase is in Fig. \ref{fgeomot}. At this position, the electric phase of the alternating current in the circled half-coil is $-180\degree$; being the phase $\theta$ defined by $I=I_m\cos(\omega t + \theta)$, where $I$ is the current, $I_m$ is the current amplitude, and $\omega$ is the angular frequency. The positive sign of the current is defined in the positive $z$ direction. The initial phase in each coil (made of two half-coils) increases by 120$\degree$ by moving clockwise.

\begin{figure}[tbp]
\centering
{\includegraphics[trim=5 84 9 58,clip,width=9 cm]{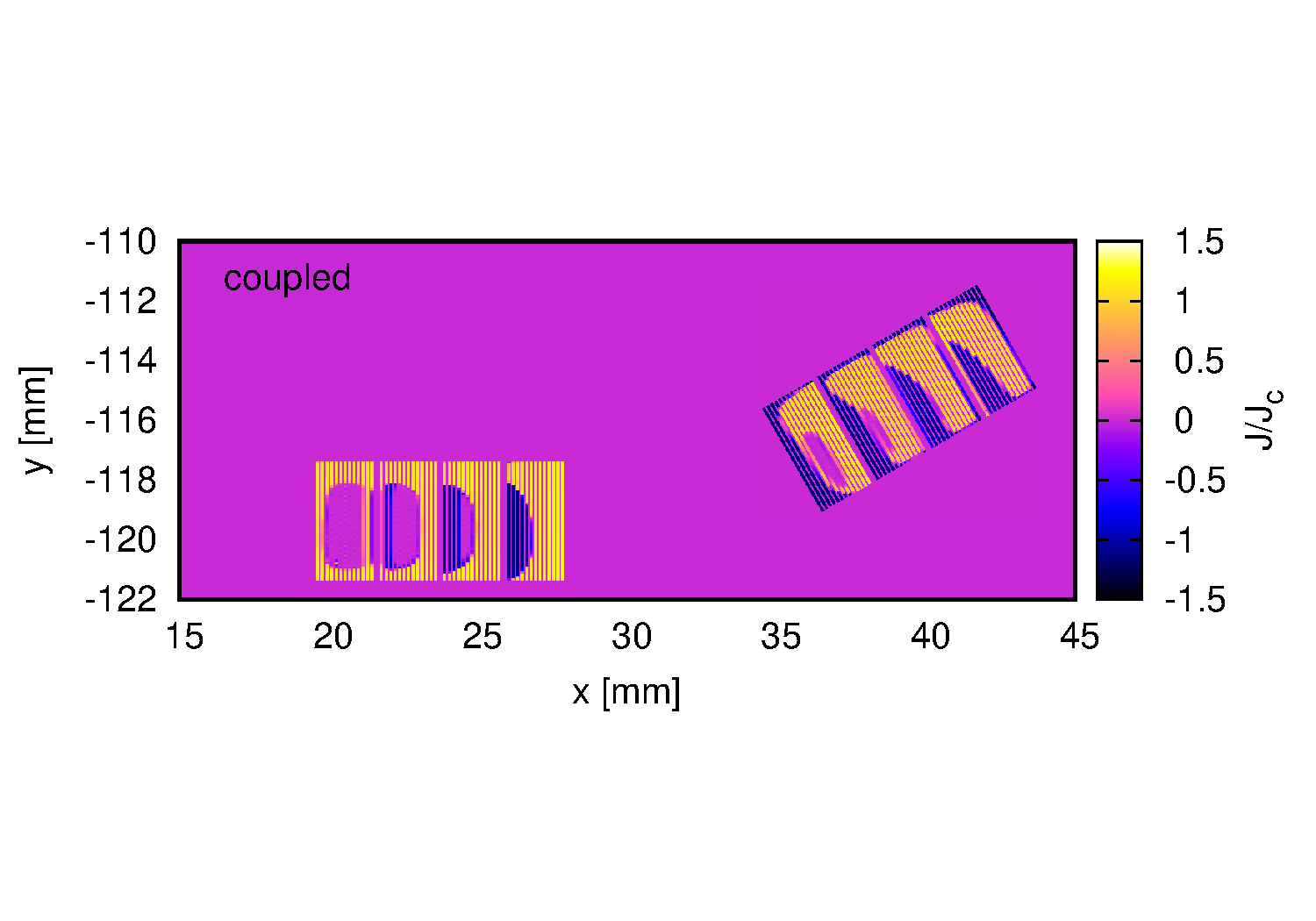}} \\
{\includegraphics[trim=5 84 9 58,clip,width=9 cm]{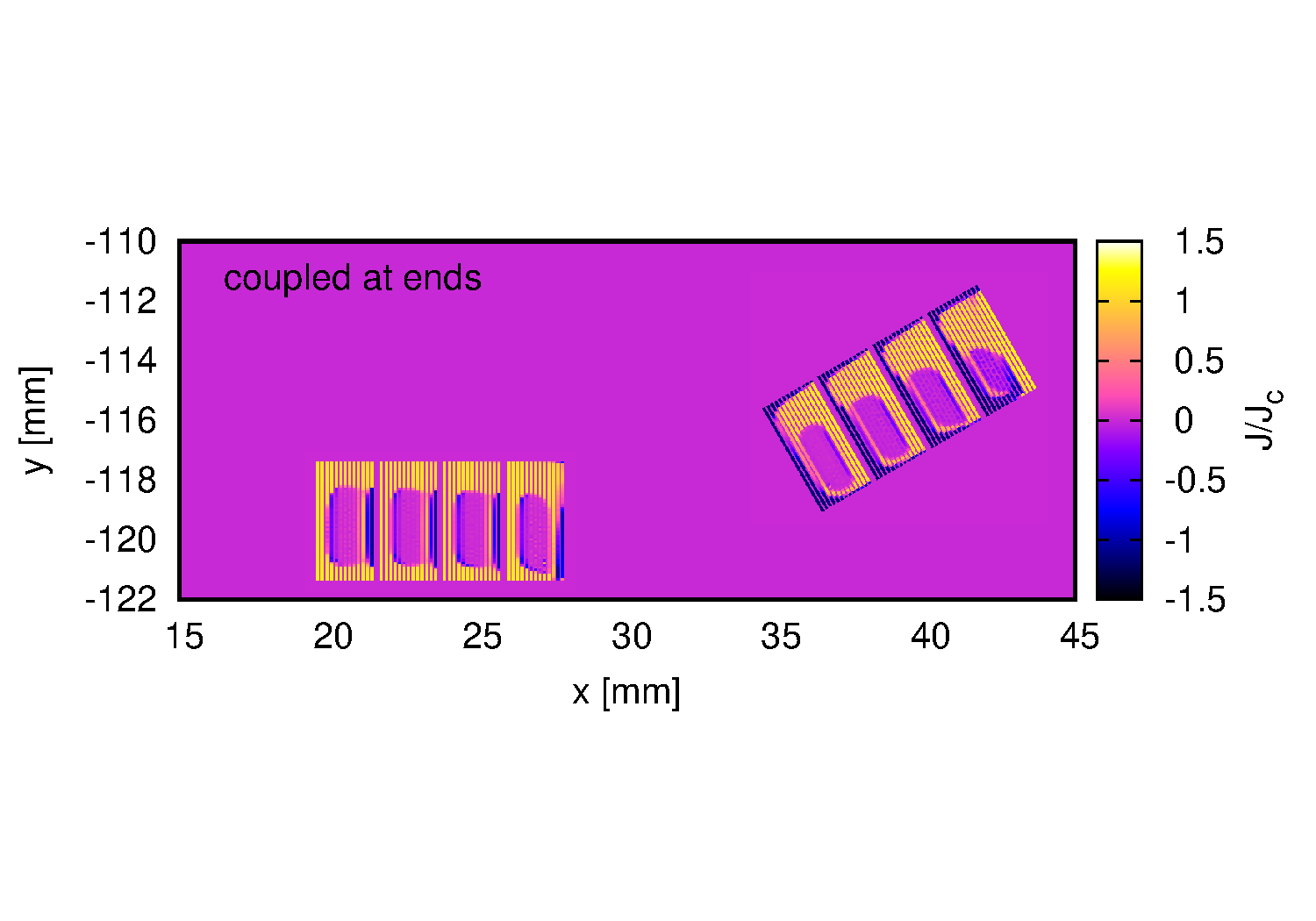}} \\
{\includegraphics[trim=5 55 9 58,clip,width=9 cm]{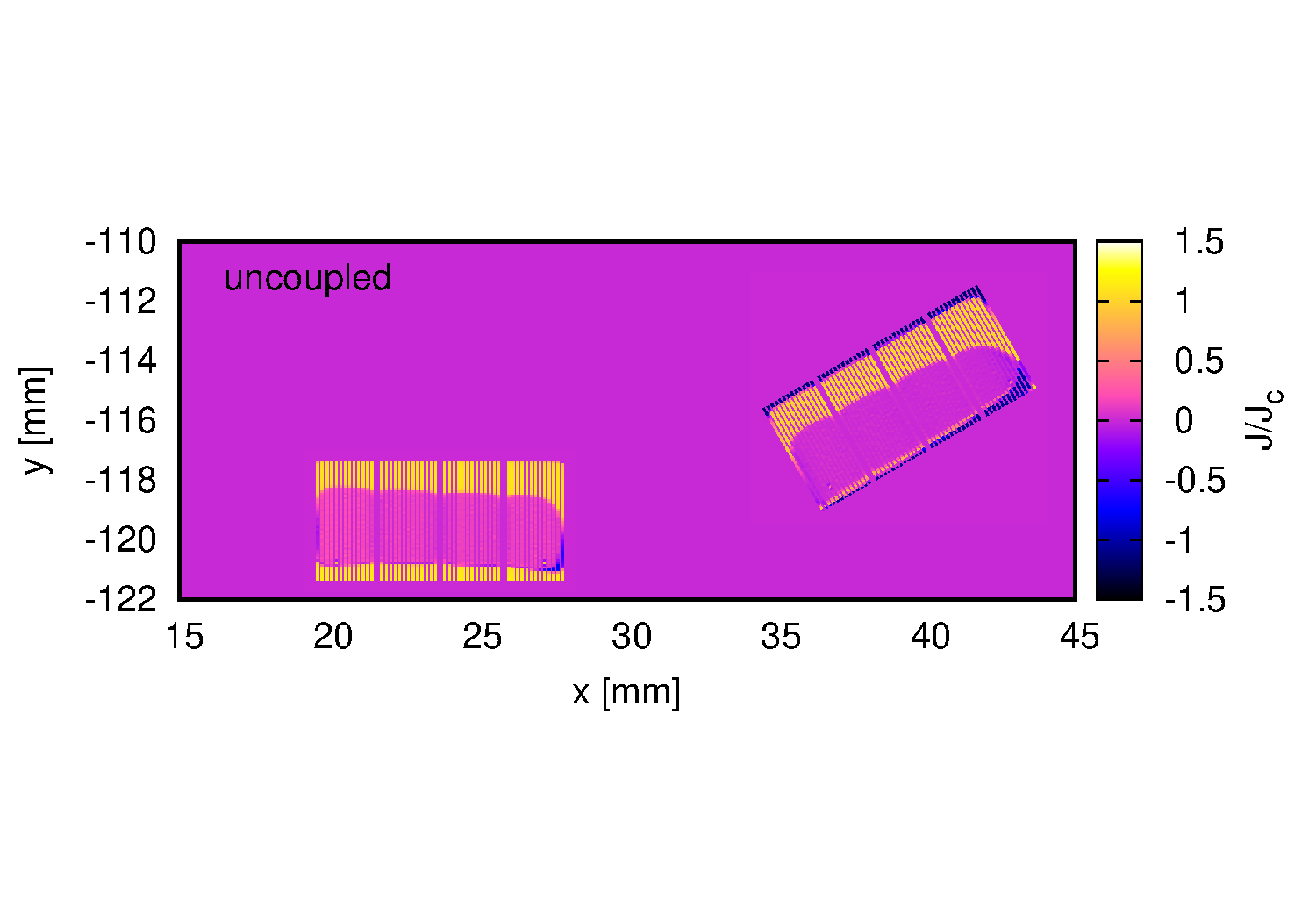}} 
\caption{Current distribution in the circled and framed half-coils of Fig. \ref{fgeomot} at the peak of the current of the circled half-coil {(left coil in this figure)}. Case for tape $I_c=$ 91.9 A.}
\label{fJmot}
\end{figure}

Again, the current density penetration is the largest for the coupled case, showing large portions with current opposite to the transport current. The current penetration is smaller for the coupled-at-ends situation and the smallest for the uncoupled case (Fig. \ref{fJmot}).

The AC loss for the coupled-at-ends case lays in between the coupled and uncoupled limits. Then, isolating the tapes along the length substantially reduces the AC loss compared to the fully coupled case. The ratio between the coupled-at-ends and uncoupled {configurations} is higher than for the single coil. The reason is the following. For the coil, the magnetic field in each half-section is practically anti-symmetric with respect to its center. Then, the net magnetic flux between tapes in the same cable due to the parallel component of the applied field practically vanishes.

\begin{figure}[tbp]
\centering
{\includegraphics[trim=0 0 0 0,clip,width=9 cm]{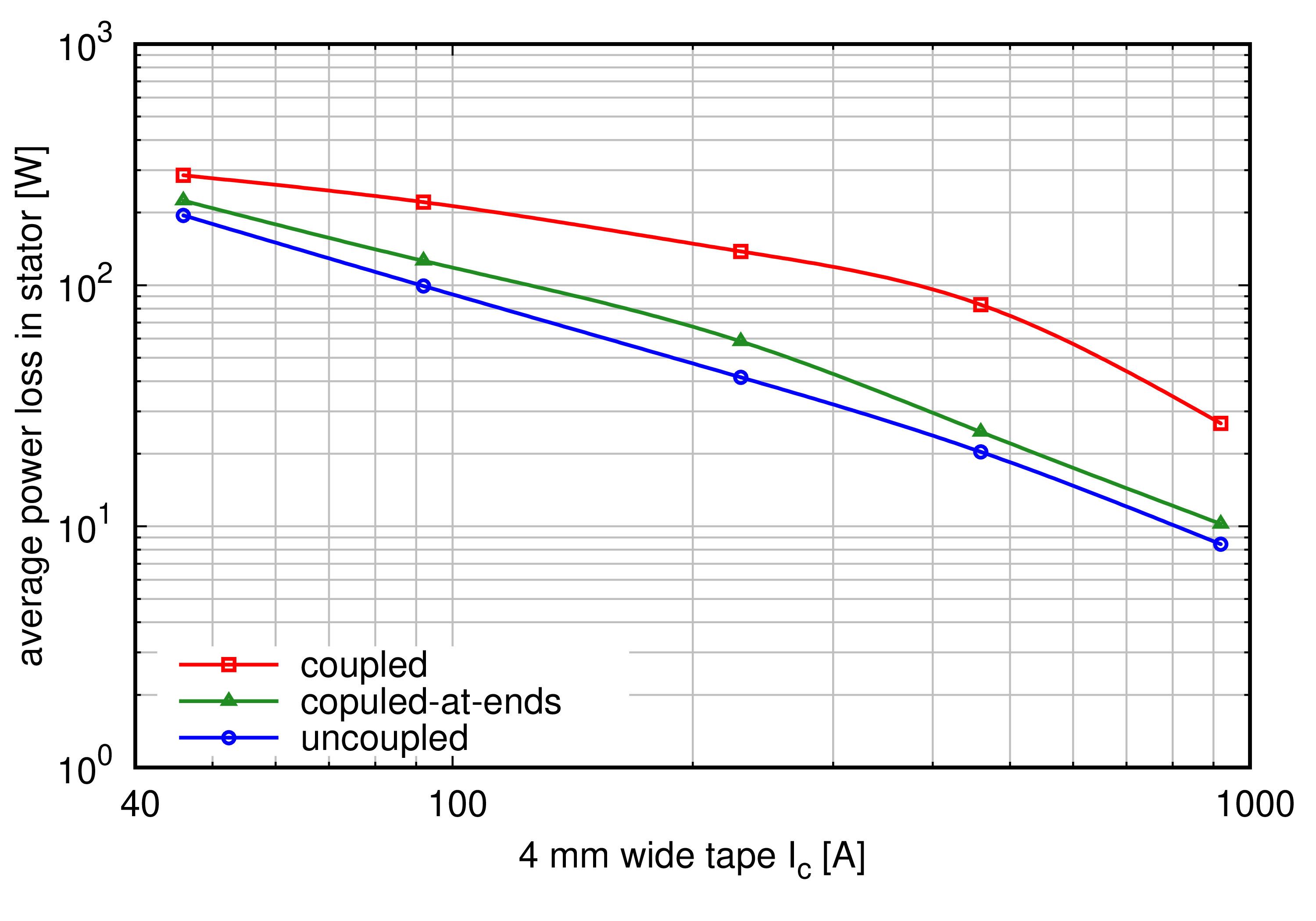}} 
\caption{Isolating the tapes along the length (coupled-at-ends situation) reduces the AC loss in the motor stator without the need of transposition.}
\label{fQmot}
\end{figure}

The plot of AC loss as a function of tape $I_c$ is useful to predict the qualitative effect of reducing the operation temperature {or improving the material pinning properties, both increasing} $I_c$. Then, Fig. \ref{fQmot} shows that the AC loss increases with temperature. Taking into account that the penalizing factor decreases with temperature, there will be an optimum operation point.

\section{Conclusion}

We developed a numerical method to compute the AC loss in the stator of superconducting motors, which combines MEMEP \cite{pardo15SST} with conventional FEM. This method could also be used for superconducting rotors and fully superconducting motors, as long as the interaction between non-linear eddy currents in the rotor with those in the stator is not important.

We modelled both single coils and a motor with windings made of multiple parallel tapes. For this, we assume three coupling scenarios: coupled, uncoupled and coupled-at-ends. 

We found that for both single coils and motor stators, isolating the tapes along their length substantially reduces the AC loss (coupled-at-ends situation).

The simplicity of the coupled-at-ends situation can be advantageous compared to the uncoupled case, which consists of a more complicated cable or power system. For example, a 4 mm wide ROEBEL cable could further reduce of the AC loss thanks to transposition, which reduces parallel field loss, and narrower strand width, which reduces perpendicular field loss. However, the crossed strands could increase the AC loss for high tape $I_c$. In addition, the cost of a ROEBEL cable is substantially higher than that of parallel tapes. Nevertheless, transposed cables with current sharing offer improved stability compared to parallel tapes.

This article shows that there are still many unexploited strategies to reduce the AC loss in rotating machines that do not involve complex transposed cables. The numerical method presented in this article will be used to explore AC loss reduction strategies for an experimental full-superconducting motor for aviation \cite{ASuMED}.








\begin{thebibliography}{10}

\bibitem{wolsky13IEA}
A.~M. Wolsky.
\newblock {HTS} from precommercial to commercial: A roadmap to future use of
  {HTS} by the power sector.
\newblock {\em International Energy Agency}, 2013.
\newblock Available at \url{http://superconductivityiea.rse-web.it}.

\bibitem{martucciello15SST}
N.~Martucciello, F.~Giubileo, G.~Grimaldi, and V.~Corato.
\newblock Introduction to the focus on superconductivity for energy.
\newblock {\em Supercond. Sci. Technol.}, 28(7):070201, 2015.

\bibitem{gieras09PET}
J.F. Gieras.
\newblock Superconducting electrical machines {--} state of the art.
\newblock {\em Przeglad Elektrotechniczny}, 85(12):1--19, 2009.

\bibitem{haran17SST}
K.~S. Haran, S.~Kalsi, T.~Arndt, H.~Karmaker, R.~Badcock, B.~Buckley,
  T.~Haugan, M.~Izumi, D.~Loder, J.~W. Bray, P.~Masson, and E.~W. Stautner.
\newblock High power density superconducting rotating machines-{Development}
  status and technology roadmap.
\newblock {\em Supercond. Sci. Technol.}, 30(12):123002, 2017.

\bibitem{ASuMED}
Advanced superconducting motor experimental demonstrator ({ASuMED}).
\newblock \url{http://www.asumed.oswald.de/}.

\bibitem{delhaye15talk}
J.L Delhaye and P.~Rostek.
\newblock Hybrid electric propulsion.
\newblock {\em Europe-Japan Symposium of Electrical Technologies for the
  Aviation of the Future}, 2015.
\newblock Airbus, avilable at
  \url{https://sunjet-project.eu/sites/default/files/Airbus%20-%20Delhaye.pdf}.

\bibitem{acreview}
F.~Grilli, E.~Pardo, A.~Stenvall, D.~N. Nguyen, W.~Yuan, and F.~G{\"om\"o}ry.
\newblock Computation of losses in {HTS} under the action of varying magnetic
  fields and currents.
\newblock {\em IEEE Trans. Appl. Supercond.}, 24(1):8200433, 2014.

\bibitem{roebelreview}
W.~Goldacker, F.~Grilli, E.~Pardo, A.~Kario, S.~I. Schlachter, and
  M.~Vojen{\v{c}}iak.
\newblock Roebel cables from {REBCO} coated conductors: a one-century-old
  concept for the superconductivity of the future.
\newblock {\em Supercond. Sci. Technol.}, 27(9):093001, 2014.

\bibitem{weiss17SST}
J.~D. Weiss, T.~Mulder, H.~J. ten Kate, and D.~C. van~der Laan.
\newblock Introduction of {CORC}{\textregistered} wires: highly flexible, round
  high-temperature superconducting wires for magnet and power transmission
  applications.
\newblock {\em Supercond. Sci. Technol.}, 30(1):014002, 2017.

\bibitem{vanderlaan18SST}
D.~C. van~der Laan, J.~D. Weiss, C.~H. Kim, L.~Graber, and S.~Pamidi.
\newblock Development of {CORC}{\textregistered} cables for helium gas cooled
  power transmission and fault current limiting applications.
\newblock {\em Supercond. Sci. Technol.}, 31(8):085011, 2018.

\bibitem{vojenciak15SST}
M.~Vojen{\v{c}}iak, A.~Kario, B.~Ringsdorf, R.~Nast, D.~C. van~der Laan,
  J.~Scheiter, A.~Jung, B.~Runtsch, F.~G{\"o}m{\"o}ry, and W.~Goldacker.
\newblock Magnetization ac loss reduction in {HTS CORC}{\textregistered} cables
  made of striated coated conductors.
\newblock {\em Supercond. Sci. Technol.}, 28(10):104006, 2015.

\bibitem{souc17SST}
J.~{\v{S}}ouc, F.~G{\"o}m{\"o}ry, M.~Vojen{\v{c}}iak, E.~Seiler,
  J.~Kov{\'a}{\v{c}}, and L.~Frolek.
\newblock Superconducting {HTS} coil made from round cable cooled by liquid
  nitrogen flow.
\newblock {\em Supercond. Sci. Technol.}, 30(10):105014, 2017.

\bibitem{takayasu16IES}
M.~Takayasu, L.~Chiesa, N.~C. Allen, and J.~V. Minervini.
\newblock Present status and recent developments of twisted stacked-tape cable
  {(TSTC)} conductor.
\newblock In {\em IEEE Trans. Appl. Supercond.}, volume~26, 2016.

\bibitem{bykovsky16IES}
N.~Bykovsky, D.~Uglietti, R.~Wesche, and P.~Bruzzone.
\newblock Design optimization of round strands made by twisted stacks of {HTS}
  tapes.
\newblock {\em IEEE Trans. Appl. Supercond.}, 26(2):4201207, 2016.

\bibitem{amemiya18SST}
N.~Amemiya, N.~Tominaga, R.~Toyomoto, T.~Nishimoto, Y.~Sogabe, S.~Yamano, and
  H.~Sakamoto.
\newblock Coupling time constants of striated and copper-plated coated
  conductors and the potential of striation to reduce shielding-current-induced
  fields in pancake coils.
\newblock {\em Supercond. Sci. Technol.}, 31(2):025007, 2018.

\bibitem{grilli07SST}
F.~Grilli and S.~P. Ashworth.
\newblock Measuring transport ac losses in {YBCO-}coated conductor coils.
\newblock {\em Supercond. Sci. Technol.}, 20:794, 2007.

\bibitem{clem07SSTb}
J.R. Clem, J.H. Claassen, and Y.~Mawatari.
\newblock {AC} losses in a finite {Z} stack using an anisotropic
  homogeneous-medium approximation.
\newblock {\em Supercond. Sci. Technol.}, 20:1130, 2007.

\bibitem{pancaketheo}
E.~Pardo.
\newblock Modeling of coated conductor pancake coils with a large number of
  turns.
\newblock {\em Supercond. Sci. Technol.}, 21:065014, 2008.

\bibitem{pardo12SSTb}
E.~Pardo, J.~{\v{S}}ouc, and J.~Kov{\'a}{\v{c}}.
\newblock {AC} loss in {ReBCO} pancake coils and stacks of them: modelling and
  measurement.
\newblock {\em Supercond. Sci. Technol.}, 25:035003, 2012.

\bibitem{zermeno11IES}
VM~Rodriguez-Zermeno, N.~Mijatovic, C.~Traeholt, T.~Zirngibl, E.~Seiler,
  AB~Abrahamsen, NF~Pedersen, and MP~Sorensen.
\newblock Towards faster {FEM} simulation of thin film superconductors: A
  multiscale approach.
\newblock {\em IEEE Trans. Appl. Supercond.}, 21(3):3273--3276, 2011.

\bibitem{ainslie11SST}
M.D. Ainslie, V.M. Rodriguez-Zermeno, Z.~Hong, W.~Yuan, T.J. Flack, and T.A.
  Coombs.
\newblock An improved {FEM} model for computing transport {AC} loss in coils
  made of {RABiTS} {YBCO} coated conductors for electric machines.
\newblock {\em Supercond. Sci. Technol.}, 24:045005, 2011.

\bibitem{grinenko12SST}
V.~Grinenko, G.~Fuchs, K.~Nenkov, C.~Stiehler, M.~Vojen{\v{c}}iak, T.~Reis,
  B.~Oswald, and B.~Holzapfel.
\newblock Transport {AC} losses of {YBCO} pancake coils wound from parallel
  connected tapes.
\newblock {\em Supercond. Sci. Technol.}, 25(7):075006, 2012.

\bibitem{grinenko13SST}
V.~Grinenko, G.~Fuchs, K.~Nenkov, and B.~Holzapfel.
\newblock An efficient method for {AC} loss reduction of {YBCO} pancake coils
  wound from parallel tapes.
\newblock {\em Supercond. Sci. Technol.}, 26(3):035002, 2013.

\bibitem{oswald12PhP}
B~Oswald, ATAM de~Waele, M~S{\"o}ll, T~Reis, T~Maier, J~Oswald,
  J~Teigelk{\"o}tter, and T~Kowalski.
\newblock Project sutor superconducting speed-controlled torque motor for
  25.000 nm.
\newblock {\em Physics Procedia}, 36:765--770, 2012.

\bibitem{motor17EUCAS}
E.~Pardo, F.~Grilli, T.~Reis, and S.~Wolfst\"adter.
\newblock Ac loss in {REBCO} stator windings of superconducting motors for
  electric and hybrid aircrafts.
\newblock {\em European Conference on Applied Superocnductivity (EUCAS)}, pages
  1LP3--04, 2017.
\newblock Available at
  \url{https://indico.cern.ch/event/659554/contributions/2714658/}.

\bibitem{motor18HTSmod}
E.~Pardo, M.~Kapolka, F.~Grilli, and T.~Reis.
\newblock Multi-physics variational methods for magnet and power applications.
\newblock {\em International Workshop on Numerical Modelling of HTS}, 2018.

\bibitem{transformer}
E.~Pardo, M.~Staines, Z.~Jiang, and N.~Glasson.
\newblock {AC} loss modelling and measurement of superconducting transformers
  with coated-conductor {Roebel-cable} in low-voltage winding.
\newblock {\em Supercond. Sci. Technol.}, 28:114008, 2015.

\bibitem{pardo15SST}
E.~Pardo, J.~{\v Souc}, and L.~{Frolek}.
\newblock Electromagnetic modelling of superconductors with a smooth
  current-voltage relation: variational principle and coils from a few turns to
  large magnets.
\newblock {\em Supercond. Sci. Technol.}, 28:044003, 2015.

\bibitem{motorTA}
T.~Benkel, Y.~Liu, E.~Pardo, S.~Wolfst\"adter, T.~Reis, and F.~Grilli.
\newblock {T-A} formulation to model electrical machines with {HTS} coated
  conductor coils.
\newblock 2019.
\newblock arXiv:1901.02370, available at
  \url{https://arxiv.org/abs/1901.02370}.

\end{thebibliography}

\end{document}